\documentclass[fp]{jpsj3}
\usepackage{txfonts}

\usepackage{multirow}
\usepackage{url}
\usepackage{color}

\title{Study on Origins of Reverse Lane Usage using Nagel-Schreckenberg Model}

\author{
Toshihiro Noda$^1$\thanks{t-noda@ai.is.saga-u.ac.jp},
Yasuhiro Hieida$^2$, and 
Shin-ichi Tadaki$^1$\thanks{tadaki@cc.saga-u.ac.jp}
}
\inst{
$^1$Department of Information Science, Saga University, Saga 840-8502, Japan \\
$^2$Computer and Network Center, Saga University, Saga 840-8502, Japan
} 

\abst{
On highways with two lanes, cars are requested to run on the slow lane and are allowed to run on the fast lane for overtaking.    
However, on real two-lane highways, we daily observe a phenomenon called ``reverse lane usage", in which the flow on the fast lane exceeds that on the slow lane.    
The origins of the phenomenon have not been discussed clearly.    
In this paper, we study a two-lane extension of the Nagel-Schreckenberg model.    
We employ two types of lane-changing rules: a German and a Japanese type.    
The German type suppresses overtaking slow cars on the fast lane through the slow lane.    
The Japanese type, on the contrary, allows overtaking through both lanes.    
If two lanes allow the same maximum speed, the suppression of overtaking through the slow lane is the key for the reverse lane usage.    
If the maximum speed on the fast lane is faster than that on the slow lane, the reverse lane usage is observed in the Japanese type.    
The effects of stochasticity in lane changing, and mixing trucks and passenger cars are also discussed.
}

\begin{document}
\maketitle
\section{Introduction}\label{chap:introduction}
Congestion in vehicle traffic is one of familiar phenomena observed in highways and city streets.    
Congestion phenomena are also observed in flows of various granular materials such as pedestrians and logistics, and have been attracting research interests.    
Since 1990's traffic flow have been studied in a physics point of view through theoretical and simulation methods.    \cite{ref:ns,ref:2,ref:3,ref:4,ref:5,ref:6,ref:7,ref:8}
In particular, experimental studies have been conducted recently and phase transition from free flow to congestion has been confirmed.    \cite{ref:9,ref:10,ref:11,ref:12,ref:13}

\begin{figure}[ht]
\centering
\includegraphics[width=5.3cm,angle=-90]{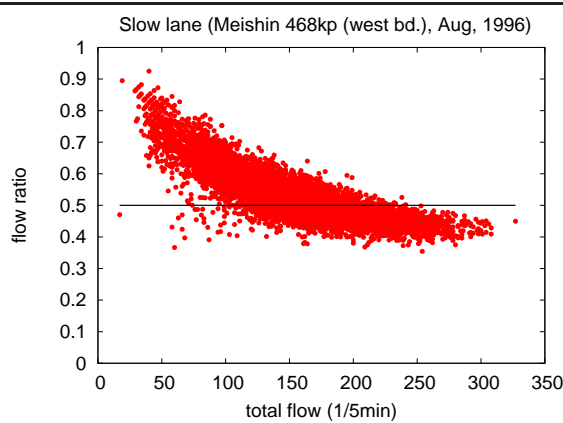}
\vspace{20pt}
\caption{
Reverse lane usage observed on the Meishin highway in August 1996.
The horizontal axis is the total flow per 5 minutes, and the vertical axis is the ratio of the flow on the slow lane to the total flow.
The flow on the fast lane exceeds that on the slow lane for large values of the total flow.
}
\label{fig:measureData}
\end{figure}

Highways usually have multiple lanes, where some interesting phenomena different from those on single-lane roads are observed.  
Figure~\ref{fig:measureData} shows the ratio of the flow on the slow lane to the total flow observed on a Japanese two-lane highway.
As the total flow increases, the ratio becomes less than the half.
In other words, the flow on the fast lane exceeds that on the slow lane, though the Japanese traffic law requires cars to run on the slow lane and allows cars to run on the fast lane only for overtaking.
We call this phenomenon as ``reverse lane usage".
The same phenomenon has been reported also in Germany.  \cite{ref:germanytwolane}

Two lane extentions of traffic flow models have been studied.
Nagatani extended the simplest traffic flow model, Wolfram's rule-184 cellular automaton, for constructing a two lane model\cite{Nagatani:1994}.
Rickert et al.\ introduced lane-changing rules into the Nagel-Schreckenberg  (NS) cellular automaton model\cite{Rickert:1996}.
Mixed flow of slow and fast cars in two lane roads has been studied based on the NS model\cite{Chowdhury:1997} and the Fukui-Ishibashi model\cite{Moussa:2003}.
Nagel et al.\ surveyed varieties of lane-changing rules\cite{ref:germanytwolane}.
Two lane models have been also used for studying effects of a single slow car\cite{Knospea} and blockades\cite{Kurata}.

A few studies have focused on reverse lane usage however.
The phenomenon has been reproduced using an extended NS  model, where the lane-changing rule contains suppression of overtaking through the slow lane\cite{ref:germanytwolane}.
The phenomenon has also been reproduced using the coupled-map Optimal Velocity model, where cars can overtake through both lanes but cars on the fast lane run faster than those on the slow lane.\cite{ref:15,ref:16,ref:twolane}
Modeling lane-changing motions contains various factors, such as demand conditions and parameters.
The purpose of this paper is to clarify the origins of the reverse lane usage.

Two-lane traffic flow models usually consist of two single-lane traffic flow models and lane-change rules.
For our purpose, complicated single-lane models reproducing realistic features are not suitable, because those models generally contain many tuning parameters.
The NS model is one of the most used cellular automaton traffic flow models.
It reproduces observed features in the fundamental diagram (density-flow relations).
To this end, we employ the NS model with an extension to two-lane roads. 

\begin{table}[ht]
\centering
\caption{Demands for lane changing.  }
\begin{tabular}{c||p{45mm}c|p{45mm}c}
\hline
Lane changing & \multicolumn{2}{c|}{Japanese type} & \multicolumn{2}{c}{German type} \\
\hline \hline
\multirow{2}{*}{From the slow lane} & & & overtake & \\
\multirow{2}{*}{to the fast lane} & overtake & Eq.~(\ref{needfast}) & or & Eq.~(\ref{needfastG}) \\
 & & & suppress overtaking from the slow lane & \\
\hline
\multirow{2}{*}{From the fast lane} & require not keeping running on the fast lane & & require not keeping running on the fast lane & \\
\multirow{2}{*}{to the slow lane}  & or & Eq.~(\ref{needslowJ}) & and & Eq.~(\ref{needslowG}) \\
 & overtake & & suppress overtaking from the slow lane & \\
\hline
\end{tabular}
\label{table:rule}
\end{table}

We study effects of two types of lane-changing rules: a German and a Japanese type.
The German type suppresses overtaking slow cars on the fast lane through the slow lane, and the Japanese one allows overtaking through both lanes.
For comparison, we construct a common part of the lane-changing rule and add characteristic conditions for the German and the Japanese type.
For simplicity, lane-changing motions are assumed to be deterministic.

This paper is organized as follows.  
Section~2 overviews the Nagel-Schreckenberg model for single-lane roads.  
In Sect.~3, we extend the model to two-lane roads by introducing a lane-changing rule.
The common part for the German and the Japanese type is discussed.
The reverse lane usage is investigated using the Japanese type of the lane-changing rule in Sect.~4.
Effects of the German type of the lane-changing rule is investigated in Sect.~5.
In Sect.~6, we investigate cases of mixed traffic with trucks (slow cars) and passenger cars (fast cars).
Effects of stochastic lane changing are discussed in Sect.~7.
Section~8 is devoted to summary and discussion.

\section{The Nagel-Schreckenberg model}\label{chap:nsmodel}
The Nagel-Schreckenberg (NS) model is one of the most used cellular automaton traffic flow models\cite{ref:ns}.
The road is divided into cells, and only one car can occupy one cell.
Each car accelerates and decelerates under the maximum speed $v_\mathrm{max}$ depending on the number of empty cells ahead.

Let us consider the $n$-th car.
The position, the speed and the distance to the preceding car are denoted by $x_n$, $v_n$, and  $\Delta x_n=x_{n+1}-x_n$, respectively.
The speed is updated by the following three steps:
\begin{description}
\item[Acceleration:]Accelerate by 1 in the range up to the maximum speed $v_\mathrm{max}$,
\begin{equation}
v_n=\min\left(v_n+1,v_\mathrm{max}\right).
\label{accele}
\end{equation}
\item[Deceleration:]Decelerate so as not to collide with the preceding car,
\begin{equation}
v_n=\min\left(v_n,\Delta x_n-1\right).
\label{slowdown}
\end{equation}
\item[Randomization:]Decelerate by 1 with probability $p$,
\begin{equation}
v_n=\max\left(v_n-1,0\right).
\label{randombreak}
\end{equation}
\end{description}

After the speeds of all cars are updated, cars run with the new value of speeds.
Thus the positions are updated virtually in parallel.

\section{Two-lane NS model}\label{chap:twolane}

In this section, we will construct a two-lane NS model by introducing a lane-changing rule shown below.
After having attempted to change lanes, all the cars move according to the single-lane NS model.

Lane-changing rules consist of a demand and a safety condition.
The demand has two types: lane changing for overtaking ($\mathrm{D}_\mathrm{O}$) and for returning to the slow lane ($\mathrm{D}_\mathrm{R}$).
The demand ($\mathrm{D}_\mathrm{O}$) is created when a car catches up its preceding car and can run faster by lane changing.
The demand ($\mathrm{D}_\mathrm{R}$) is created when there is no need to run on the fast lane.
The safety condition (S) is that a car avoids being collided by cars running on its adjacent lane if the car moves to the adjacent lane.
If a car has the demand, the car moves to the target lane only if the safety condition (S) is satisfied.

On two-lane highways, in general, cars are requested to run on the slow lane and are allowed to use the fast lane for overtaking.
For constructing lane-changing rules, it is necessary to consider suppression of overtaking through the slow lane.
The Japanese traffic law requires to use the fast lane only for overtaking\cite{ref:douro}.
However, there is no instruction for suppressing overtaking slow cars on the fast lane.
On the contrary, the German traffic law has the instruction \cite{ref:germanytwolane}.
If a car running on the slow lane finds a slow car running on the fast lane, the car is requested to move behind the slow car on the fast lane for giving room for the slow car going back to the slow lane.

We are interested in the origins of the reverse lane usage.
We investigate effects of the Japanese and the German type of lane changing.
The German type of lane-changing rules in this paper is basically the same as those in Ref.\citen{ref:germanytwolane}.
For comparison, the common part of lane-changing rules among the two types is constructed first.
Next, characteristic parts of the two types are added to the common part.

Table~\ref{table:rule} summarizes the demand conditions in the Japanese and the German type of lane changing.  
For lane changing from the slow lane to the fast lane (the upper row in Table~\ref{table:rule}), a condition for overtaking ($\mathrm{D}_\mathrm{O}$) the preceding car on the slow lane is common for both types.  
As for the German type, there is another condition for not overtaking a slow car on the fast lane.  
For lane changing from the fast lane to the slow lane (the lower row in Table~\ref{table:rule}), there is a common condition to require cars to return to the slow lane in the case of no need to keep running on the fast lane ($\mathrm{D}_\mathrm{R}$).    
The Japanese type has an additional condition for overtaking ($\mathrm{D}_\mathrm{O}$) a slow car on the fast lane.  
And the German type has an additional condition to prevent cars from overtaking a slow car on the fast lane  through the slow lane.  

\begin{figure}[ht]
\centering
\includegraphics[width=8.5cm]{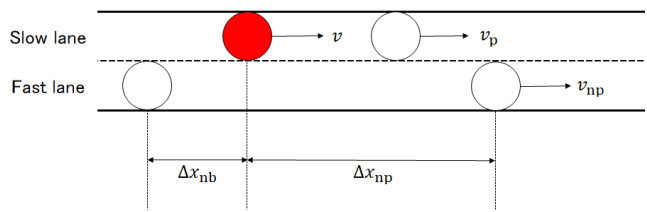}
\caption{
A conceptual diagram of lane changing from the slow lane to the fast lane.  
The circles represent cars.  
The red circle denotes the car focused on.  
}
\label{fig:concept}
\end{figure}

Figure~\ref{fig:concept} is a conceptual diagram of lane changing from the
slow lane to the fast lane.  
Let us focus our attention on the car represented by the red circle in Fig.~\ref{fig:concept}.    
The speed of the car is $v$ and the speed of its preceding car running on the same lane is $v_\mathrm{p}$.    
On the adjacent lane, there is another preceding car running with speed $v_\mathrm{np}$, whom the headway distance to is $\Delta x_\mathrm{np}$.    
There is another car running $\Delta x_\mathrm{nb}$ behind on the adjacent lane.  

First we construct the common parts of the condition which generates the demand $\mathrm{D}_\mathrm{O}$ for overtaking by lane changing.  
For lane changing from the slow lane to the fast lane, the common condition is for overtaking the preceding car on the slow lane (see the upper row in Table~\ref{table:rule}).    
It consists of that the preceding car on the same (slow) lane runs slower than the car focused on, and that the preceding car on the adjacent (fast) lane runs faster than the preceding car on the same lane,
\begin{equation}
v_{\mathrm{p}}\leq v \mathrm{\ AND\ } v_{\mathrm{p}}<v_{\mathrm{np}}.    
\label{needfast}
\end{equation}

For lane changing from the fast lane to the slow lane, the common condition is to request for returning to the slow lane in the case of  no need to keep running on the fast lane ($\mathrm{D}_\mathrm{R}$, see the lower row in Table~\ref{table:rule}) .    
It is that the speed of the preceding car on the adjacent lane is faster than that of the car focused on,
\begin{equation}
v_{\mathrm{np}}>v.  
\label{needslow}
\end{equation}

If the preceding cars are distant enough, drivers do not care about the speed of those preceding cars.  
For this effect, we introduce the {\it distance of vision} $d$.
If a preceding car on any lanes is distant over $d$, the speed of it is set infinite.  
In other word, drivers measure the speed of the preceding cars only within the distance $d$.    
The distance of vision $d$ is mainly set to 16.  
We remark that $d=16$ corresponds to about $100 \mathrm{m}$ if $v_\mathrm{max}\simeq 120 \mathrm{km/h}$.

Next we construct the safety condition (S) for lane changing.  
This condition is common to the two types.  
For avoiding collision, the following condition must be satisfied,
\begin{equation}
\Delta x_\mathrm{np}>v \mathrm{\ AND\ } \Delta x_\mathrm{nb}>v^\ast_\mathrm{max},
\label{safety}
\end{equation}
where $v^\ast_{\mathrm{max}}$ denotes the maximum speed of the car running behind on the adjacent lane seen from the focused car.   
In all sections except Sect.~\ref{chap:defferent_vmax},
\begin{itemize}
\item $v^\ast_\mathrm{max}=v^{\mathrm{f}}_{\mathrm{max}}$ in the case of motions from the slow lane to the fast lane, and
\item $v^\ast_\mathrm{max}=v^{\mathrm{s}}_{\mathrm{max}}$ in the case of motions from the fast lane to the slow lane.  
\end{itemize}

At every time step, all cars are selected in random order for lane changing.  
If a car has the demand for lane changing and the safety condition is satisfied, the car moves to the adjacent lane immediately.  
After all cars are given chances for lane changing, they move according to the single-lane NS model.  

In the following sections, the simulation system has 10,000 cells for each lane under periodic boundary conditions.  
For each simulation, cars are randomly distributed on both lanes initially.  
After the relaxation with 1,000 time steps, we start measurement.  
The maximum speed on the slow lane $v_\mathrm{max}^\mathrm{s}$ is set to 5.  
In all simulations, we observe the ratio of the flow on the slow lane to the total flow, and determine whether the reverse lane usage occurs or not with several sequences of random numbers.  

Real highways are definitely open-boundary systems. 
Moreover those roads are very inhomogeneous with ramps, curves, sags, and so on.
Some of such inhomogeneous factors work as bottlenecks and induce traffic jams.
For inducing traffic jams in simulation systems with open boundaries, you need introduce bottlenecks or strong noises.
In any cases, traffic states depend on observation points.
In viewpoint of our purpose, we employ periodic boundaries for simplicity.

\section{Japanese two-lane model}\label{chap:ruleJ}

\begin{figure*}[ht]
\centering
\includegraphics[width=8cm]{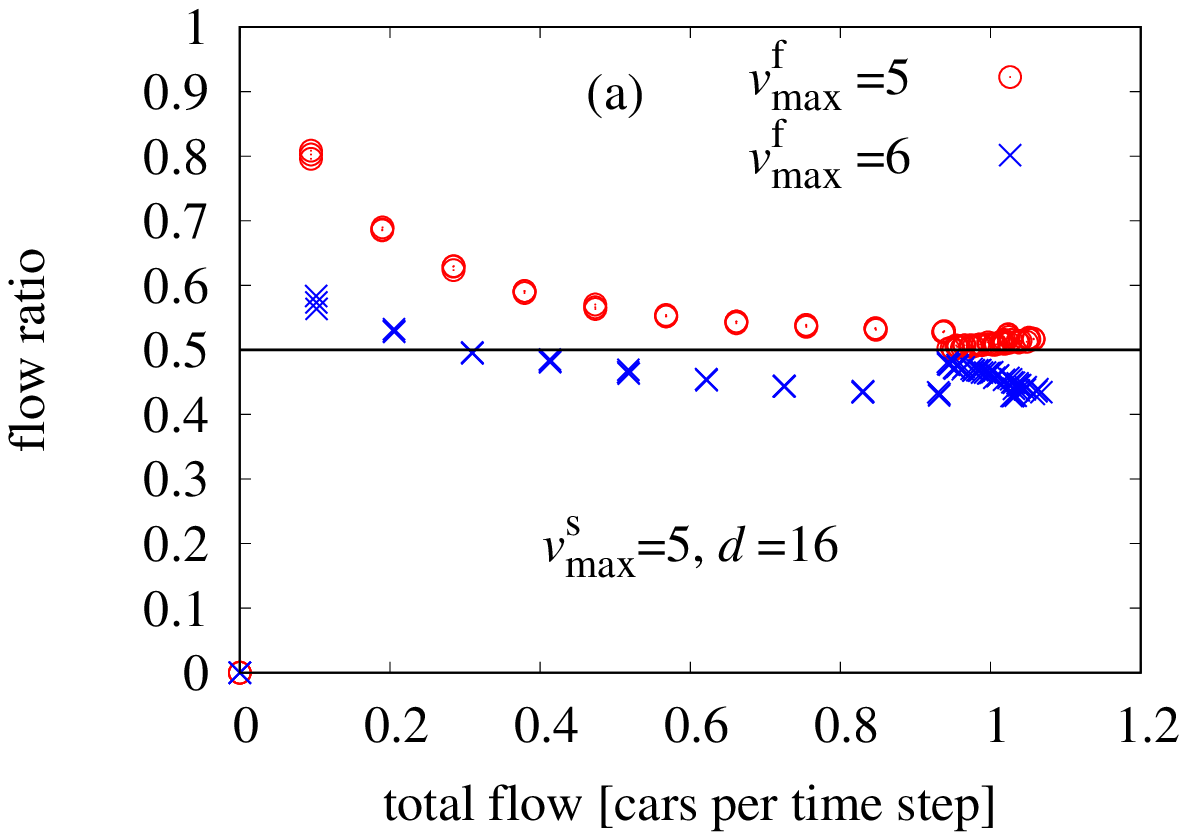}
\includegraphics[width=8cm]{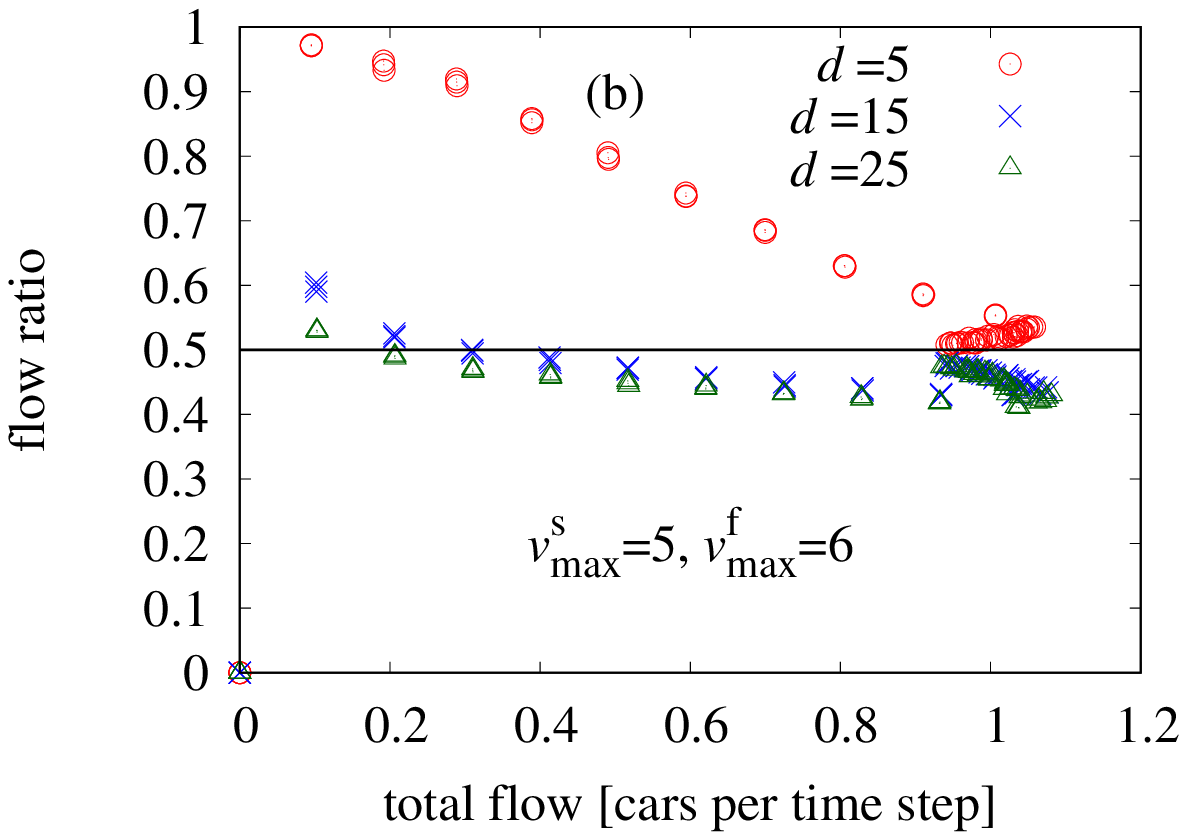}
\vspace{10mm}
\caption{
Simulation results of the ratio of the flow on the slow lane to the total flow in the Japanese type of the lane-changing rule.  
Figure (a) compares results of different values of the maximum speed of the fast lane $v^\mathrm{f}_\mathrm{max} = 5$ and $v^\mathrm{f}_\mathrm{max} = 6$.    
For $v^\mathrm{f}_\mathrm{max} = 6$, the reverse lane usage occurs.  
Figure (b) shows the effect of the distance of vision $d$.    
If $d$ is short, the reverse lane usage does not occur.  
}
\label{fig:resultJ}
\end{figure*}

We investigate the relation between the Japanese type of the lane-changing rule and the reverse lane usage in this section.  
In the Japanese type, as mentioned previously (Table~\ref{table:rule}), the demand for moving from the slow lane to the fast lane aims only to overtake the preceding slow car and its condition is expressed by Eq.~(\ref{needfast}).    
On the other hand, the demand for moving from the fast lane to the slow lane consists of two parts: the common demand ($\mathrm{D}_\mathrm{R}$) for requiring returning to the slow lane and the characteristic demand ($\mathrm{D}_\mathrm{O}$) for overtaking the slow preceding car on the fast lane.  
The condition for the demand ($\mathrm{D}_\mathrm{R}$) is the common one (Eq.(\ref{needslow})) which is in the first half ($v_\mathrm{np}>v$) of Eq.(\ref{needslowJ}).    
The condition for the demand ($\mathrm{D}_\mathrm{O}$) is the second half of Eq.(\ref{needslowJ}): 
\begin{equation}
v_{\mathrm{np}}>v \mathrm{\ OR\ } (v_{\mathrm{p}}\leq v \mathrm{\ AND\ } v_{\mathrm{p}}<v_{\mathrm{np}}).    
\label{needslowJ}
\end{equation}
The safety condition consists of only the common part represented by Eq.~(\ref{safety}).

The maximum speed on the fast lane $v_\mathrm{max}^\mathrm{f}$ varies as a parameter.  
Figure~\ref{fig:resultJ}(a) shows the simulation results for two different values of $v_\mathrm{max}^\mathrm{f}$.    
If the maximum speed on the fast lane equals that on the slow lane, $v_\mathrm{max}^\mathrm{f}=v_\mathrm{max}^\mathrm{s}$, the reverse lane usage does not occur.  
On the other hand, the reverse lane usage occurs for the case $v_\mathrm{max}^\mathrm{f}>v_\mathrm{max}^\mathrm{s}$.    
In other words, the reverse lane usage is the effect where cars on the fast lane runs faster than those on the slow lane.

Figure~\ref{fig:resultJ}(b) shows the effects of the distance of vision $d$, where we set $v_\mathrm{max}^\mathrm{f}>v_\mathrm{max}^\mathrm{s}$.    
If the distance of vision is set short, $d=v_\mathrm{max}^\mathrm{s}=5$, the reverse lane usage does not occur.  
This is understood as follows.  
Let us think a car running at the maximum speed.  
It runs and finally finds the slow preceding car at the distance $d$.    
For lane changing, the safety condition must be satisfied.  
The safety condition requires $2\times v^\ast_\mathrm{max}$ empty sites on the adjacent lane and consequently is not frequently satisfied.  
If the car cannot change the lane, the car must decelerate to follow the slow preceding car.  
In this way, cars lose opportunities for lane changing if $d$ is short.  
For $d$ is long ($d=15$ and $25$), the reverse lane usage occurs, because the opportunities for lane changing is large.  
And the longer the distance of vision is, the greater the opportunities to change lanes are.  

\section{German two-lane model}\label{chap:ruleG}

\begin{figure*}[ht]
\centering
\includegraphics[width=8cm]{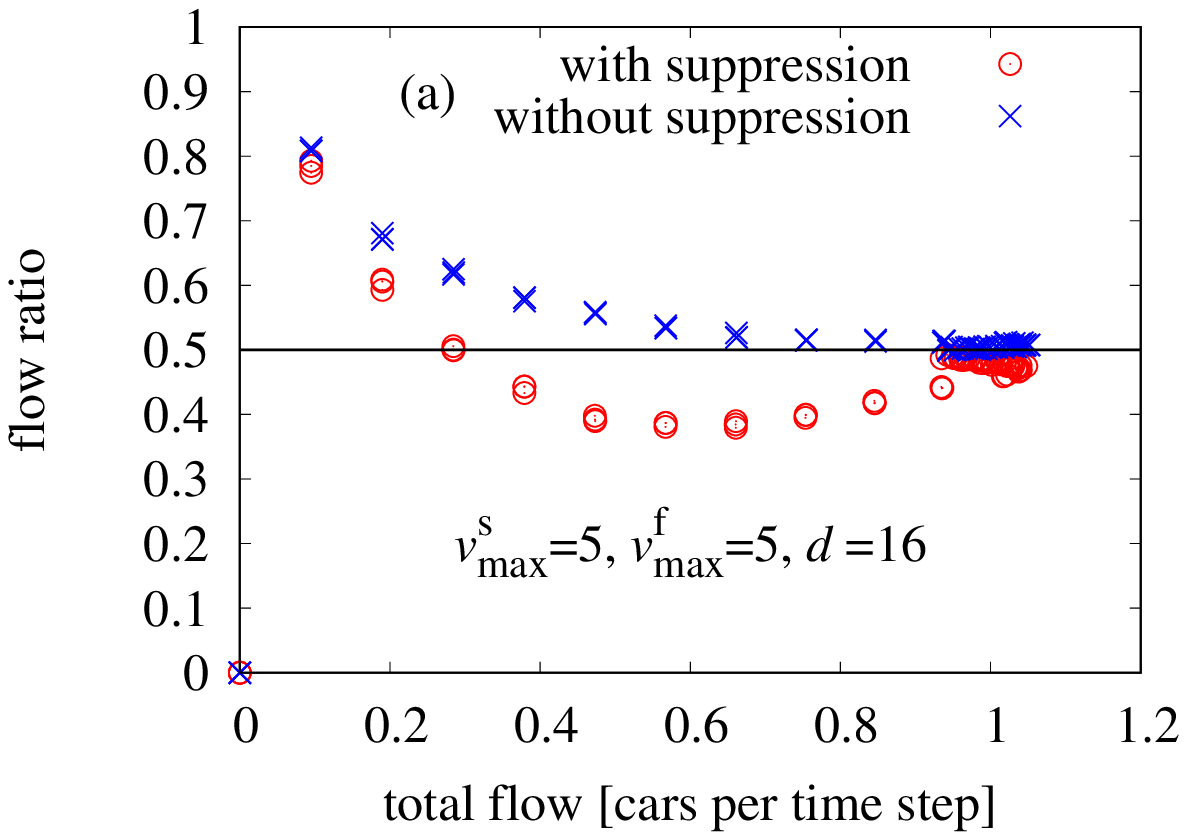}
\includegraphics[width=8cm]{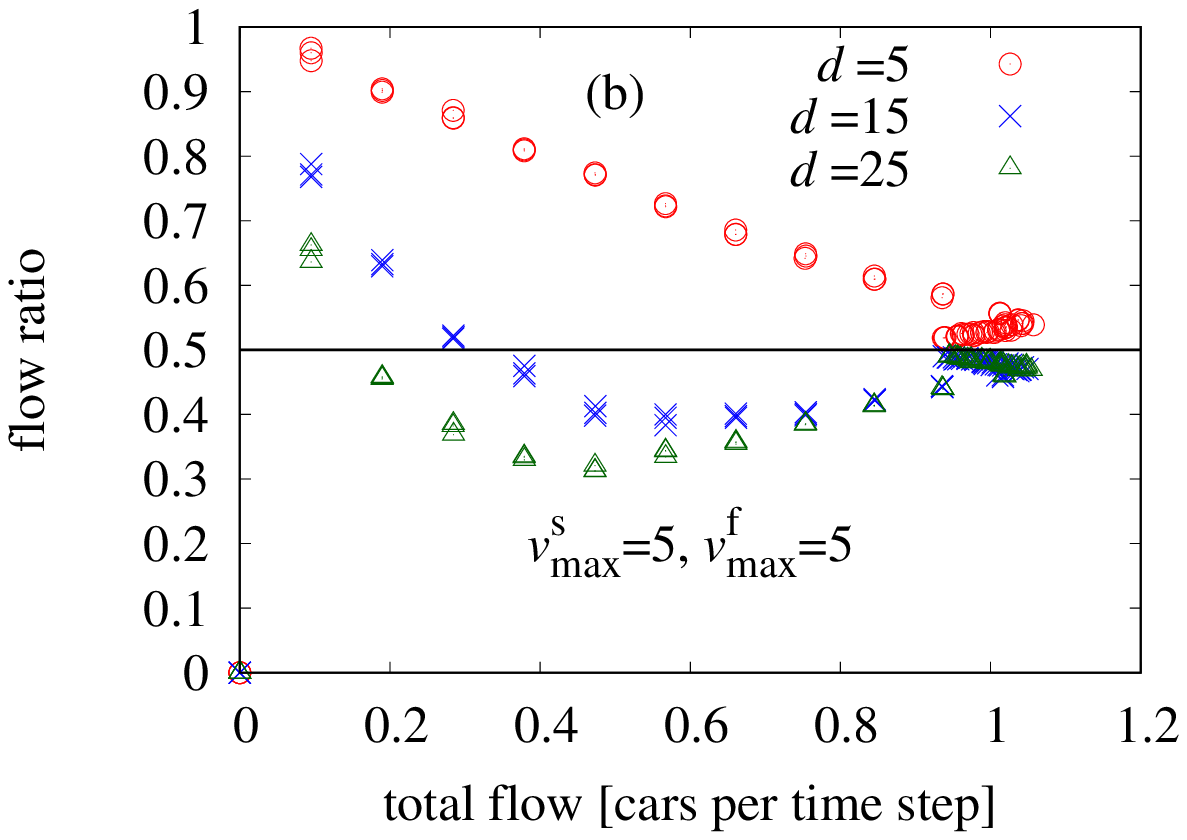}
\vspace{1.0cm}
\caption{
Simulation results of the ratio of the flow on the slow lane to the total flow in the German type of the lane-changing rule.  
Figure (a) compares results with and without the suppression of overtaking from the slow lane.  
The reverse lane usage occurs only with the suppression.  
Figure (b) shows results with different values of the distance of vision $d$.    
The suppression of overtaking from the slow lane is included.  
The reverse lane usage does not occur for $d=5$.    
The longer the distance of vision is, the more the flow on the fast lane is.  
}
\label{fig:resultG}
\end{figure*}

Next, we investigate the relation between the German type of the lane-changing rule and the reverse lane usage.  
The characteristic of the German type is that overtaking slow cars on the fast lane through the slow lane is suppressed.  
To this end, a car is requested to move to the fast lane if the car finds a slow car on the fast lane.  
This motion creates enough space where the slow car can return safely to the slow lane.  
Additionally, the lane changing from the fast lane to the slow lane is also suppressed when a car finds the slow preceding car on the fast lane.  
If the car changes the lane, the car ends up overtaking the slow preceding car from the slow lane. 

In the condition of the demand moving from the slow lane to the fast lane, the suppression of overtaking through the slow lane is added to the common condition (Eq.~(\ref{needfast})) as the last part ($v_\mathrm{np}\le v$),
\begin{equation}
(v_{\mathrm{p}}\leq v \mathrm{\ AND\ } v_{\mathrm{p}}<v_{\mathrm{np}}) \mathrm{\ OR\ } v_{\mathrm{np}}\leq v.  
\label{needfastG}
\end{equation}
If a car runs on the fast lane, in addition to the common condition (Eq.~(\ref{needslow})) for requiring returning to the slow lane, a new condition ($v_\mathrm{p}>v$) for suppressing overtaking the slow preceding car on the fast lane through the slow lane is required,
\begin{equation}
v_{\mathrm{np}}>v \mathrm{\ AND\ } v_{\mathrm{p}}>v.  
\label{needslowG}
\end{equation}
Even if the slow lane is clear, a car cannot move to the slow lane if the preceding car is slower than itself.  
The car must give room for the preceding car to return to the slow lane.  
As in the Japanese type, the safety condition consists of only the common part represented by Eq.~(\ref{safety}).

For observing the effects of the suppression of overtaking through the slow lane, we perform two types of simulations: with and without the suppression.  
In the system without the suppression, the condition of the demand for lane changing consists of only the common parts.  
For this purpose, we set the same maximum speed for cars on both lanes.

Figure~\ref{fig:resultG}(a) shows the simulation results for the cases with and without the suppression of overtaking through the slow lane.  
In the case of the latter, the reverse lane usage does not occur.  
In other words, the suppression of overtaking through the slow lane is important for the reverse lane usage.

\begin{figure*}[ht]
\centering
\includegraphics[width=8cm]{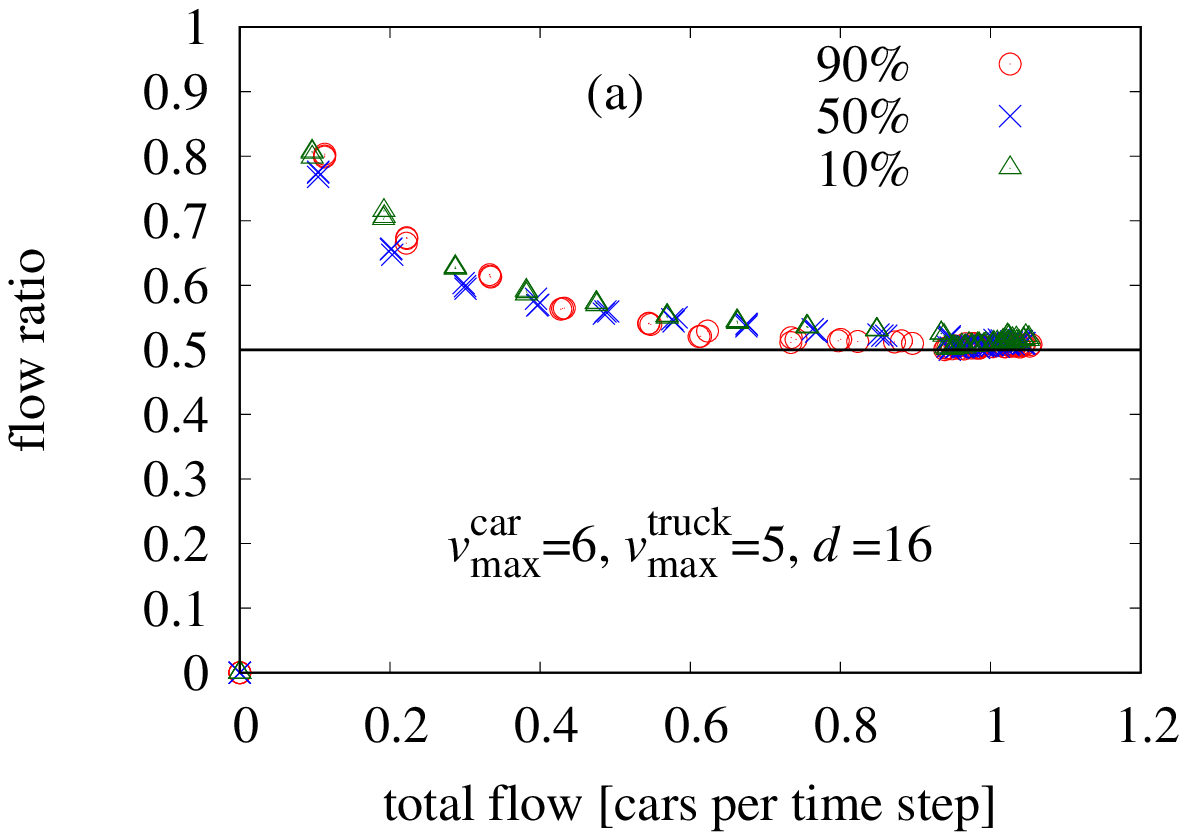}
\includegraphics[width=8cm]{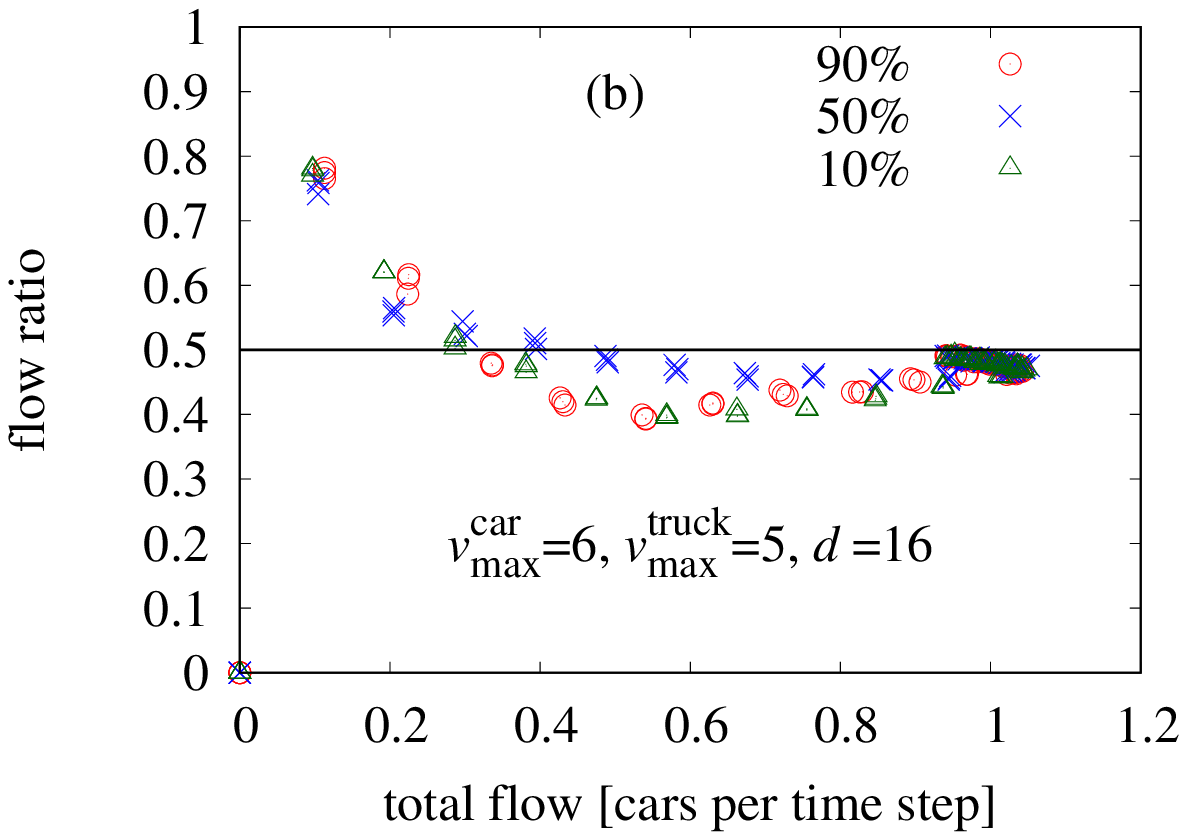}
\vspace{1.0cm}
\caption{
Simulation results of the ratio of the flow on the slow lane to the total flow in mixed traffic.  
Figure (a) shows the results for the Japanese type of the lane-changing rule.  
There are no significant difference between different mixing ratios.  
Figure (b) shows the results for the German type of the lane-changing rule.  
All cases show the reverse lane usage.  
The mixing ratio affects the flow ratio.  
}
\label{fig:resultD}
\end{figure*}

\begin{figure*}[ht]
\centering
\includegraphics[width=8cm]{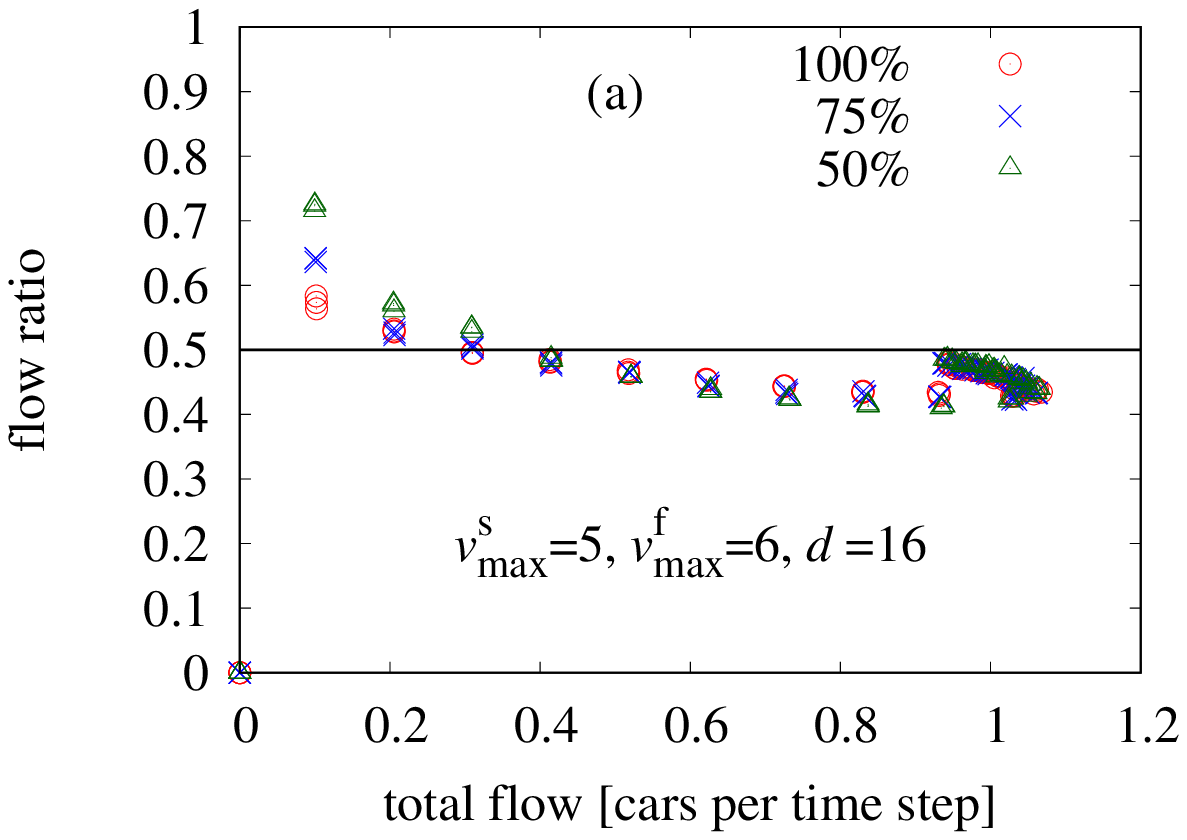}
\includegraphics[width=8cm]{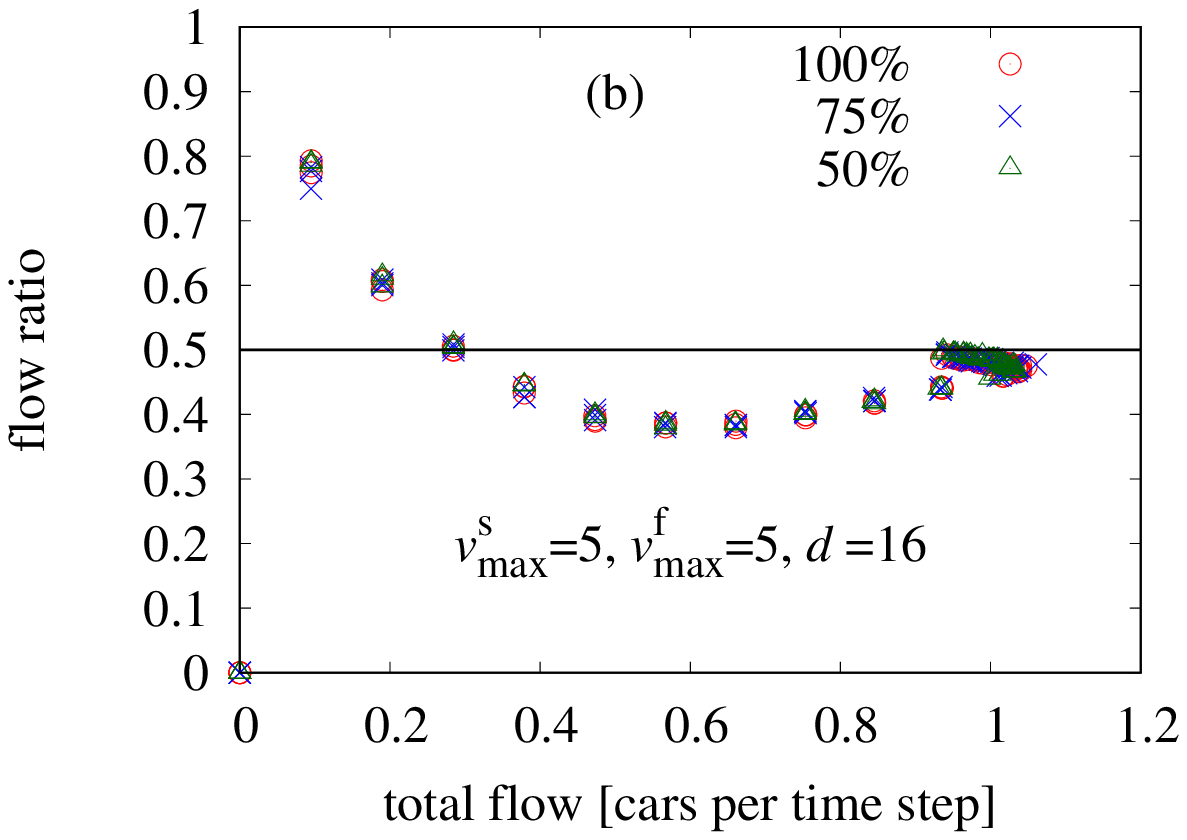}
\vspace{1.0cm}
\caption{
Simulation results of the ratio of the flow on the slow lane to the total flow in the case of stochastic lane changing.  
Results for the Japanese and the German type of lane changing are shown in Figure (a) and (b), respectively.  
For the Japanese type, the maximum speeds are set $v^\mathrm{s}_\mathrm{max}=5$ for the slow lane and $v^\mathrm{f}_\mathrm{max}=6$ for the fast lane.  
The German type includes the suppression of overtaking through the slow lane.  
The stochasticity does not affect the results significantly.  
}
\label{fig:resultP}
\end{figure*}

For the effects of the distance of vision $d$, we find, in Fig.~\ref{fig:resultG}(b), the same effects as in the Japanese type.  
If the distance of vision is short, $d=5$, the reverse lane usage does not occur.  
The longer the distance of vision is, the greater the chances to change lanes are.  

\section{Mixed traffic of trucks and passenger cars}\label{chap:defferent_vmax}

In this section, we investigate the reverse lane usage in mixed traffic of trucks (slow cars) and passenger cars (fast cars), where the maximum speed of each car depends only on whether the car is a truck or a passenger car.   
In this section, $v^\ast_\mathrm{max}$ in the safety condition (Eq.~(\ref{safety})) denotes the maximum speed of the car running behind on the adjacent lane seen from the focused car.  

For both the Japanese and the German type, we set the maximum speeds of passenger cars to $v^\mathrm{car}_\mathrm{max}=6$ and those of trucks to $v^\mathrm{truck}_\mathrm{max}=5$.    
The ratio of passenger cars to all cars is set to $90\%$, $50\%$ and $10\%$.

Figure~\ref{fig:resultD}(a) shows simulation results for the Japanese type.  
For all values of the ratio of passenger cars, the reverse lane usage does not occur.  
We may expect that the reverse lane usage will occur by running trucks on the slow lane and passenger cars on the fast lane.   
Contrary to this naive expectation, it is guessed that passenger cars do not concentrate on the fast lane and trucks not on the slow lane.  

Figure~\ref{fig:resultD}(b) shows simulation results for the German type.  
In all cases of the ratio of passenger cars, the reverse lane usage occurs.  
In the case of $50\%$ passenger cars, the flow ratio is higher than other cases, which means the flow on the fast lane exceeds a little that on the slow lane.  
Anyway, the inhomogeneity of car speed suppresses the reverse lane usage for the German type.  

\section{Stochastic lane changing}

We have assumed, so far for simplicity, that a car moves to the adjacent lane whenever the car has the demand for lane changing and satisfies the safety condition.  
In real highways, of course, cars move to the adjacent lane stochastically even if the conditions are fulfilled.  
In this section, we study the effects of stochasticity in lane changing on the reverse lane usage.   

In the following, we set the probability $p_\mathrm{L}$ of lane changing equal for both motions, from the slow lane to the fast one and from the fast lane to the slow one.  
The probability $p_\mathrm{L}$ is set to $100\%$, $75\%$ and $50\%$.    
For the Japanese type, the maximum speed is $v^\mathrm{s}_\mathrm{max}=5$ on  the slow lane and $v^\mathrm{f}_\mathrm{max}=6$ on the fast lane.  
For the German type, the maximum speed is $v_\mathrm{max}=5$ on the both lanes.

Simulation results are shown in Fig.~\ref{fig:resultP}(a) for the Japanese type and Fig.~\ref{fig:resultP}(b) for the German type.  
All cases show the occurrence of the reverse lane usage.  
And we also find that the stochasticity in lane changing does not affect the results significantly.  

\section{Summary and discussion}
In this paper, we have studied the reverse lane usage using an extended Nagel-Schreckenberg traffic flow model.  
We have implemented the Japanese and the German type of lane changing.  
For comparison, these two types share the common part.  

The Japanese type of lane changing allows overtaking through both lanes.  
The reverse lane usage only occurs when the maximum speed of cars on the fast lane is faster than that on the slow lane.  
On the contrary, the German type of lane changing suppresses overtaking through the slow lane.  
The reverse lane usage occurs even if the maximum speed of cars on the fast lane equals that on the slow lane.  

We have introduced the distance of vision $d$ within which a car observes the speeds of the preceding cars on both lanes.  
We have observed that, in both the Japanese and the German type, if $d$ is short (resp.   long), the reverse lane usage does not (resp.   does) occur.  
This is because the longer the distance of vision is, the greater the chances to change lanes are.  

The effects of inhomogeneity of speed have been studied.  
We have prepared mixed traffic of trucks (slow cars) and passenger cars (fast cars).    
By naive expectation, it is guessed that passenger cars may use the fast lane and trucks do the slow lane, and that the reverse lane usage may occur even in the case of equal maximum speed on both lane for the Japanese type.  
However, it does not occur.  
Mixed traffic does not affect the results significantly.  

We also have found that the stochasticity in lane changing does not affect the results significantly.

Mixed traffic of trucks and passenger cars was studied in Ref.~\citen{Chowdhury:1997}.
The lane-changing demands are different from our study.
Their \textit{symmetric} lane-changing rules are for overtaking a preceding car if the gap to the preceding car is less than $1+v$, where $v$ is the speed of the car under consideration.
Their \textit{asymmetric} lane-changing rules consist of two additional ones to the symmetric ones.
The first one are applied only for passenger cars on the slow lane, where passenger cars move to the fast lane if they find a truck in front.
The second one are only for trucks on the fast lane, where trucks return back to the slow lane if the slow lane is clear.
There are no rules for suppressing overtaking slow cars through the slow lane.
In this sense, their rules have effects of enhancing passenger cars using the fast lane and suppressing trucks using the fast lane.
As a result, passenger cars prefer running on the fast lane and trucks on the slow lane.

We employed the Nagel-Schreckenberg model for the single-lane traffic flow. 
The model is simple and has been used for simulations.
The model, however, does not show, for instance, metastability and hysteresis observed in real traffic and reproduced by other traffic flow models\cite{TakayasuTakayasu:1993,Benjamin:1996,Barlovic:1998}.
One of the common features in these models showing metastability is slow-to-start motion where stopped cars delay restarting.
This feature affects mainly the macroscopic states in high density regions.
The reverse lane usage is observed, on the contrary, in intermediate density regions.
Our preliminary simulation results with the model in Ref.~\citen{Benjamin:1996}) reproduce the qualitatively same results as those using the NS model.

\bibliography{reference}
\end{document}